# Staring Down the Digital Fulda Gap – Path Dependency as a Cyber Defense Vulnerability

Jan Kallberg[1]

*[1]Army Cyber Institute at West Point, United States Military Academy, West Point, New York, United States*

Jan Kallberg, Army Cyber Institute at West Point, United States Military Academy, Spellman Hall 4-33, 2101 New South Post Road, West Point, NY 10996, United States
E-mail: jkallberg@gwu.edu

## Staring Down the Digital Fulda Gap – Path Dependency as a Cyber Defense Vulnerability

**Introduction**

Academia, homeland security, defense, and media have accepted the perception that critical infrastructure in a future cyber war (cyber conflict) is the main gateway for a massive cyber assault on the U.S. The question is not if the assumption is correct or not; the question is instead of how did we arrive at that assumption. The cyber paradigm considers critical infrastructure the primary attack vector for future cyber conflicts. The national vulnerability embedded in critical infrastructure is given a position in the cyber discourse as close to an unquestionable truth – as a natural law.

The operational environment changes, but organizations adapt to these changes if the organizational ideology can accept the changes aligned with the dominant organizational perception of reality.[1] Large organizations develop a narrative, a story, that defines the perception of the mission, why the organization exists, and how it justifies its existence and share of shared resources. In government, bureaucratic maximization is common, where public agencies seek to maximize their size and maintain growth by adding tasks and redefining the mission (charter).[2]

History provides numerous examples of how organizations preferred to see the operational environment based on the organizational ideology and narrative. The first battles between the Germans and French armies in the First World War were examples of how military bias and organizational ideology came to play as both sides had designed their strategies based on path-dependency, bias, and organizational ideology. Both the Germans and the French had chosen offensive strategies; one of the reasons was the inherited notion that only professional armies could attack[3] and, therefore, defensive strategies were considered to undermine the army's status and professional standing. If there were a defensive doctrine, conscripted units would receive an equal standing, which the professional army did not prefer. So even if it was evident after the Boer War and the Russo-Japanese War of 1904-1905 that increased firepower, industrial mass fabricated weaponry, and the deployment of machine guns at grand scale had tipped the scale in favor of a defensive posture,[4] both the French and German armies internal bias[5] was stronger than properly assessing reality. The misalignment between perception, doctrine, and reality led to the massive loss of human lives in the First World War.

Critical infrastructure fits well the internal bias among different stakeholders; homeland security needs a significant domestic cyber threat to justify its cyber engagement, the military struggle to visualize what a cyber conflict entails, and critical infrastructure serves as an understandable target for the general public. The military influence in the cyber security discussion directly influences the perception by utilizing terminology similar to military actions. If there is a threat to critical infrastructure, the doctrine would require cyber deterrence to deter potential attackers.[6]

Any fighting force has an embedded urge to understand the battlefield, plan, strategize, and prepare for war. Historically, it has been necessary to avoid strategic surprise, horrific defeat and limit the actual casualties and devastation to its force and society. The study of the future fight has been instrumental in preventing strategic surprise. In a future cyberwar, it is at critical infrastructure where the cyber enemy will appear, or do we expect them to attack the critical infrastructure because we have decided so in over two decades of cyberwar planning and strategic discussions?

**FORGING HISTORICAL AND CYBER NARRATIVE**

Defense, homeland security or military, are large organizations formed by formal structures, hierarchies, institutional knowledge, shared history, and a common perception of reality through disseminated values that go through ranks and surface in doctrine. Beyond mission statements and quadrennial defense reviews, the defense establishment has a living narrative of who they are, what tasks are ahead, and why they are doing what we do. Traditionally, a shared view of the battlespace has been instrumental in our ability to understand war. Being aware of who you are and what you do is an asset and strength for any organization. It is the foundation for professionalism, respect, and dedication to supporting the mission. A flipside comes with a shared perspective – the risk for groupthink.[7] The large bureaucracy has invested time, money, and effort in critical infrastructure as the predominant attack vector, which adds to the path dependency as sunk cost fallacy overrides logic.[8] The unified internal view of the operational environment could become a gospel, preaching, and identifier for membership in the "professional" group as defined by the group itself.

**CRITICAL INFRASTRUCTURE AS AN INTELLECTUAL CATCH-ALL**

The assumption that the enemy in cyber will appear at a specific location, conduct pre-assessed operations, and act to predetermine the larger constraints is likely spurious because it does not reflect the operational realities of a cyber conflict.

Why are we so convinced that future massive cyber attacks will follow a specific pattern of a broad attack on critical infrastructure?

One of the unique tenets of cyber is the large target area as any networked device can be reached, the sizeable vulnerable target area is naturally inviting, but the adversary does not have endless assets to engage in their cyber mission force. The size of the U.S. economy is 21 trillion USD GDP (2021). Over 3 200 counties, 70 000+ local government departments, public works, and numerous major private utilities are already highly skilled in defensive cyber operations (DCO) visualize the size of the defending force. Even a larger adversary will have to prioritize the offensive cyber operations (OCO) targeting drastically; there are a limited number of OCO teams in the adversary's mission force.

**THE EVER INCREASING SCOPE OF CRITICAL INFRASTRUCTURE**

Since the 1980s, the definition of critical infrastructure has expanded. The definition of critical infrastructure is now so broad that it becomes a catch-all. With today's broad definition of critical infrastructure covering a spectrum starting with lightning for ballparks to e-mail alerts from career websites, it is apparent critical infrastructure's assumption faces another challenge. It is almost impossible to prioritize defensive cyber assets to defend critical infrastructure because of the broad scope.

In a 1983 report, the Congressional Budget Office (CBO) defined "infrastructure" as facilities with "the common characteristics of capital intensiveness and high public investment at all levels of government. They are, moreover, directly critical to activity in the nation's economy." The CBO included highways, public transit systems, wastewater treatment works, water resources, air traffic control, airports, and municipal water supply in this category. The CBO also noted that the concept of infrastructure could be "applied broadly to include such social facilities as schools, hospitals, and prisons, and it often includes industrial capacity, as well."[9] On Jul. 15, 1996, President Clinton signed Executive Order 13010 establishing the President's Commission on Critical Infrastructure Protection (PCCIP). This Executive Order defined "infrastructure" as The framework of interdependent networks and systems comprising identifiable industries, institutions (including people and procedures), and distribution capabilities that provide a reliable flow of products and services essential to the defense and economic security of the United States, the smooth functioning of government at all levels, and society as a whole.[10]

Since the 1990s have the cyber mantra been the protection of critical infrastructure[11] and still is. The 2013 Presidential Executive Order - Improving Critical Infrastructure Cybersecurity[12] uses a broadened definition of critical infrastructure. It is based on our previous experience, the human need to grasp and comprehend. As humans, we are colored of the past because past experiences continued through doctrine, technique, and tactics.

In May 2021, President Biden issued Executive Order 14208, "Executive Order on Improving the Nation's Cybersecurity,"[13] which reemphasizes the importance of critical infrastructure and seeks to increase threat intelligence sharing within government and between the government and the private sector.

President Biden introduced a few new innovative approaches. One of these is to create a Cyber Safety Review Board tailored after the National Transportation Safety Board that investigates airplane crashes and other significant transportation disasters. The aviation industry and the government investigates airplane crashes to find out the root cause of the incident, with the overarching goal that a vulnerability should only lead to one crash. According to the current administration, the concept of airplane crash investigations and adjustments of safety posture afterward could be transposed to cyber security. President Biden also wants to limit barriers for government and private sector cyber threat information sharing. The increasingly fast cyberattacks, from the penetration of a system to widespread effects, drive the need for information sharing.

The time window to act is becoming smaller and smaller. Information sharing of threat intelligence between the government and the private sector becomes necessary to ensure that detection and mitigation at a broad scale are not unnecessarily delayed.

**THE RATIONALE FOR CRITICAL INFRASTRUCTURE**

The increased interest in the cyber-defense of critical infrastructure has several drivers. First, our society is far more dependent on technical infrastructure than ever before, with an increasing demand for better communications, more bandwidth, and high close to undisrupted availability.[14] Forty years ago, disruptions happened, and the demand for constant infrastructural services was not present. Today's younger generation has grown up with a functional web of accessible infrastructural services, a cultural shift that increases vulnerability.

Second, in the last two decades, an increasing number of cyber-physical equipment has gone online. In the early 1990s, there was manual control of a midsized dam in a watershed. Today, in 2021, the dams in a whole watershed are connected to the Internet and controlled remotely from a control room potentially hundreds of miles away.[15] In most cases, the remotely controlled cyber-physical systems no longer have a fallback ability to use the older control methodology. The manual mode is no longer in the present system design; the earlier staff that knew how to run the system manually is no longer there. Attacks on cyber-physical equipment can lead to releasing chemicals, waste products, and other contamination that have long-term environmental damage.[16][17]

Third, along with critical infrastructures, goes supply chain concerns, as one of the outfalls of failed infrastructure is disruptions in the supply chains. The logistic chain that hauls our goods feeds our population and ensures that products safely arrive in time is increasingly exposed to cyber threats. The safe handling from farm to table that through years been manually monitored is now rapidly becoming controlled by computers.[18] Satellite communication and GPS track Refrigeration trucks' position, the cargo's temperature, and other essential data for food safety, and this data is a part of a data network through logistics, warehouses, and the supermarket.

With various operating programs and software, the number of systems coming from different technical generations, both modern and all the way to close to obsolete, create a kaleidoscope of vulnerabilities and opportunities for a potential attacker. There is no discussion that critical infrastructure is a vital component of a functional modern society; it is more a question of if critical infrastructure is unavoidably the main battlefield for a future cyber conflict.

**STRATEGIC INTENT**

Cyberattacks on critical infrastructure can have different intents. There is a similarity between cyber and national intelligence; both are trying to make sense of limited information looking at a denied information environment. In reality, our knowledge of the strategic intent and goals of our potential adversaries is limited.

We can study doctrine[19], published statements, non-classified events, and with some luck in the covert realm, the potential adversary's classified events, but still, there are significant gaps. We are assessing the adversary's strategic intent from the outside, which are often qualified guesses, with all the uncertainty that comes with it. Then to assess strategic intent, many times, logic and past behavior are the only guidance. Nation-state actors tend to seek a geopolitical end goal, change policy, destabilize the target nation, or acquire the information they can use for their benefit.

Criminal networks seek economic gain, visible in ransomware that encrypts and blocks the usage of a targeted computer network, allowing it only to be restored after the victim paid a ransom.[20]

Attacks on critical infrastructure make the news headline, and for a less able potential adversary, it can serve as a way to show their internal audience that they can threaten the United States. In 2013, Iranian hackers broke into the control system of a dam in Rye Brook, N.Y.[21] The actual damage was limited due to circumstances the hackers did not know. Maintenance procedures occurred at the facility, which limited the risk for broader damage. The limited intrusion in the control system made national news, engaged the State of New York, elected officials, Department of Justice, the Federal Bureau of Investigations, Department of Homeland Security, and several more agencies. Time Magazine called it in the headline;" Iranian Cyber Attack on New York Dam Shows Future of War." For a geopolitically inferior country that seeks to be a threat and a challenger to the U.S., examples are Iran or North Korea; the massive American reaction to a limited attack on critical infrastructure serves its purpose. The attacker had shown its domestic audience that they could shake the Americans, primarily when U.S. authorities attributed the attack to Iranian hackers, making it easier to present it as news for the Iranian audience. When attacks occur on critical domestic infrastructure, it is not a given that it has a strategic intent to damage the U.S.; the attacks can also be a message to the attacker's population that their country can strike the Americans in their homeland.

Russian leaders under sanctions can not enjoy the "benefits" of corruption as extravagant travel, moving money freely in the West, can not let their anger out except for authorizing harassing cyber-attacks on the Americans. Cyber-attacks become a risk-free way of picking a fight with the Americans without risking escalation.

Several potential adversaries do not subscribe to our playbook, especially when it comes to international law, the law of armed conflict, and ethics, so their actions are more restricted by what their senior leaders accept than anything else. Numerous cyber-attacks on critical American infrastructure could be a way to harass the American society and have no other justification than hostile authoritarian senior leaders has it as an outlet for their frustration and anger against the U.S.

Nothing stops an authoritarian regime from using American infrastructure as a training group for their cyber forces because they are not constrained by our rules, laws, and ethics. They could train their forces on our systems, with no other intent than education with the hope of gaining valuable information.

**POTENTIAL OUTCOME**

One open question is how will the American population react to a massive, crippling attack on critical infrastructure. Historically, Americans tend to react heavily to attacks on American soil with anger and support for punitive action, even if requiring a more prolonged operation. Attackers seeking to maximize civilian hardship as a tool to bring down a targeted society have historically faced a reversed reaction. The reactions to Pearl Harbor and 9-11 show that there is a risk for any adversary to attack the American homeland and that such an attack might unify American society instead of injecting fear and force submission to foreign will. The German bombings of the civilian targets during the 1940's air campaign "Blitz" only hardened the British resistance against the Nazis. An attacker needs to take into consideration the potential outfall of a significant attack on critical infrastructure.

A systematic large-scale cyberattack on the power grid, leading to a blackout similar to the Northeastern blackout of 2003,[22] does not need to lead to the same outcome as it did then. During the Northeastern blackout of 2003,[23] instead of panic, there was a sense of solidarity; New Yorkers helped each other. Rationally, there is no guarantee that the population would act in the same manner 18 years later.

The reliance on older experiences could be a blind spot as society changes over time, and new technology enables the sharing of fearful events, which could induce panic, several societal destabilizations, and second-tier effects. Will the Tik-tok, Instagram, social media population we have today be as composed and orderly as the New Yorkers in the 2003 blackout? In 2003, the population only saw their part of the event, a dark street or a stuck subway car, but it is still there. A gadget-reliant population that suddenly loses cell phone, data, light, and connectivity might react entirely differently because the impact affects their virtual and physical worlds. In the older examples, as the 2003 blackout, the virtual world was not fully developed, and the impact might have felt less. There are significant uncertainties about how the population will react to the significant cyber-enabled devastation of our infrastructure, if successful, believing that it will be as orderly as the blackout of 2003 could be wishful thinking.

**FULDA GAP**

Critical infrastructure is not the first time the U.S. had a prevailing narrative on where the main future battle would occur. The Fulda Gap analogy becomes relevant due to the decades' long assumption that if there were a Soviet Warsaw Pact sudden attack on NATO, this would be where the thrust of the Soviet Warsaw Pact attack would be. During the Cold War, NATO leaders, American and European politicians and military, and security analysts were convinced that a potential war between the Soviet Warsaw Pact and NATO would start in a flat area in central West-Germany on the border between East-Germany (DDR) and West-Germany (BRD). As a gateway to the West with heavily forested areas on both sides, the open landscape formed an invasion corridor that leads from the Soviet formations right through West Germany. In the way stood the U.S. V. Corps as the NATO defenders of the Fulda Gap sector. The Fulda Gap was by NATO the assumed attack route for the Soviet 8th Guards Army, and following Soviet and East German units, towards the Frankfurt-am-Main and Rhine, in the pursuit to have a decisive victory over NATO before U.S. troops from the continental U.S. could arrive. General James H. Polk, then commander of U.S. Army Europe and Seventh Army, stated in 1970 that:

> "...the communist armies are bigger and better trained than at any time since World War II. They have the most modern weapons and equipment....In the Soviet zone of Germany, 20 Russian tank and motorized rifle divisions.. . 300,000 men stand fully combat ready....In addition to the Soviet divisions stationed in Eastern Europe, the armies of East Germany, Czechoslovakia and Poland provide 30 additional tank and motorized rifle divisions...". [24]

The defenses against a potential Soviet rapid onslaught in central Germany dominated NATO and U.S. ground war strategic thinking for over forty years.[25] On Mar. 1, 1987, L.A. Times wrote: "NATO planners have pinpointed the Fulda Gap--several open passes running through the hills about 60 miles northeast of Frankfurt--as a likely invasion route into Western Europe for Soviet Bloc forces." The first line of defense was the U.S. Armored Cavalry; for the first half of the 40 years, the U.S. 14th Armored Cavalry Regiment was then relieved by the 11th Armored Cavalry Regiment.

The LA Times article from 1987 continued, "This is the frontier where it would happen," Col. Thomas E. White, 43, of Detroit, commander of the 4,500-man Fulda-based U.S. 11th Armored Cavalry, said in an interview." In Fulda Gap, fighting the enemy became a science.[26] The U.S. and NATO conducted studies of the target surface and exposure time of potentially invading tanks, the number of neutralized Soviet tanks per lost AH64, fuel consumption and maintenance hours for every sortie, and all summarized in impeccable statistics and scientific analysis.

The perception of the Fulda Gap as the Cold War's most crucial defense line survived the Cold War and still exists. In 2004, then-Secretary Colin L. Powell stated in a published article in the German newspaper Frankfurter Allgemeine, as a reflection on the Cold war; "Our mission was to defend the Fulda Gap, a break in the Vogelsberg mountains through which the Iron Curtain ran, and through which the Red Army any day might pour."[27]

In the current debate, the narrow space between Kaliningrad (Königsberg) and Belarus, a Russian ally, has taken over as the Fulda Gap of the 2000s.[28] The current expectation is that the narrow Lithuanian corridor that separates Kaliningrad from Belarus is the potential future battleground. The concept of Fulda Gap-thinking has survived into a new era.

**SEVEN DAYS TO THE RHINE**

After the Cold War, the Warsaw Pact war planning became accessible, especially after Poland, Czechoslovakia, Hungary, and Eastern Germany left the Soviet/Russian sphere. The Soviet Union and the Warsaw Pact made several plans through the Cold War, but they did not emphasize the Fulda Gap as NATO, the U.S., and their war planners assumed.[29] [30]

The Fulda Gap area was in the Warsaw Pact, Soviet indirectly, war planning but as one of five thrusts into the NATO defenses. According to the 1979 battle plans, the planned main Warsaw Pact/Soviet attack was further north to cut off debarkation ports to disallow British and U.S. units to land in continental Europe and secure the Baltic Sea approaches. The Warsaw Pact/Soviets made several war plans, but the Fulda Gap as the overshadowing pivot in the battle between NATO and the Soviets was a NATO, and U.S., assumption. The reason why Fulda Gap became identical to where a potential Warsaw Pact/Soviet massive armored attack could have many reasons. First, it was suitable terrain, but not the only one for a rapid assault from East to West. Second, after a breach in Fulda, the Warsaw Pact/Soviet forces were close to West Germany's industrial centers of the Rhine valley and the Ruhr, the West German then-capital Bonn, main NATO headquarters and installations. The focus on the Fulda Gap assumption fitted the DOD institutional agenda and a narrative tailored for the general public and politicians to justify programs, weapons systems and paved the way for a modernization of the U.S. Army that enabled success in the Gulf War. The narrative spread to politicians in favor of higher defense spending as Fulda Gap becomes synonymous with where the Soviet onslaught would occur if there ever were a NATO/Warsaw-pact war.

**CONCLUSION: RISKING SPURIOUS ASSUMPTIONS**

Any military organization will seek to optimize the opportunity to support its strategic goals with limited resources and assets. We tend to see vulnerabilities and concerns about cyber threats to critical infrastructure that matter to us. But an adversary will assess where and how a cyberattack will benefit the adversary's strategy. Logically, it is not convincing that attacks on general critical infrastructure have the high payoff that an adversary seeks, so it becomes questionable that the U.S. strong focus on critical infrastructure is warranted. Naturally, the U.S. will defend critical infrastructure, but if too much focus is

The American reaction to Sept. 11, and any attack on U.S. soil, hint to an adversary that attacking critical infrastructure to create hardship for the population could work contrary to the intended softening of the will to resist foreign influence. It is more likely that attacks that affect the general population instead strengthen the will to resist and fight, similar to the British reaction to the German bombing campaign "Blitzen" in 1940.

We can't rule out attacks that affect the general population, but there are not enough adversarial offensive capabilities to attack all 16 critical infrastructure sectors and gain strategic momentum. An adversary has limited cyberattack capabilities and needs to prioritize cyber targets that are aligned with the overall strategy. Logically, an adversary will focus their OCO on operations that has national security implications and support their military operations by denying, degrading, and confusing the U.S. information environment and U.S. cyber assets.